\documentclass[aps,prl,twocolumn,superscriptaddress,showpacs]{revtex4}
\usepackage{graphicx}% Include figure files

\begin{document}

\title{Unusual Length Dependence of Conductance of Some Oligomers}
\author{Y.X. Zhou}
\affiliation{Department of Physics, Fudan University, Shanghai
200433, People's Republic of China}
\author{F. Jiang}
\affiliation{Department of Physics, Fudan University, Shanghai
200433, People's Republic of China}
\author{H. Chen }
 \email[Corresponding author. Email: ] {haochen@fudan.edu.cn}
\affiliation{Department of Physics, Fudan University, Shanghai
200433, People's Republic of China}
\author{R. Note}
\affiliation{Institute for Materials Research, Tohoku University,
Sendai 980-8577, Japan}
\author{H. Mizuseki} \affiliation{Institute for Materials Research, Tohoku
University, Sendai 980-8577, Japan}
\author{Y. Kawazoe}
\affiliation{Institute for Materials Research, Tohoku University,
Sendai 980-8577, Japan}

\date{\today}

\begin{abstract}
  Recent experiment found a quantum length dependence of oligothiophene molecule conductance at low bias
  [Xu et al., nano Lett. \textbf{5}, 1491 (2005)], the long molecule has large conductance.
  By means of a first-principles method we obtain both the quantum length dependence of conductance
  at low bias and the classical length dependence of conductance at high bias region for oligothiophene.
  In between there is an oscillated conductance behavior. The transport behaviors are determined
  by the distinct electronic structures of the molecular compounds.
  The various conductance length dependence may appear for the organic compounds.
  Our further investigation finds that the classical conductance
  length dependence in polyphenanthrene dithiolates and another
  unusual conductance length dependence in polyacene ditholates:
  the quantum length dependence of conductance is at the high bias and
  the classical length dependence of conductance is at the low bias.
\end{abstract}
%73.23.-b Electronic transport in mesoscopic systems
%85.65.+h Molecular electronic devices
%31.15.Ar Ab initio calculations

\maketitle

  Recently, much interest has been focused on molecular electronics which is in expectation
  of new generation of electronic devices as so-called molecular devices. The
  fabrications of molecular devices require to understand transport
  properties of single molecules, and there are much experimental and
  theoretical works reported in these years \cite{1,2,3,4,5,6,7,8,9}.
  Some interesting effects attracted much
  attention: gate effect \cite{10,11,12}, negative differential
  resistance \cite{13,14}, and diode rectifier etc \cite{15}.
  These effects are expected as the basis of molecule device design and the
  investigation of the molecular devices are in a rapid progress. Except for
  these effects, the length dependence of conductance of molecule wires
  is also discussed widely \cite{15,16,17,18,19,20,21}.

  Lang and Avouris \cite{17} studied carbon-atom wires with the first-principles method
  and found a conductance oscillation with respect to the carbon atom number.
  Similar full ab initio computation of carbon-atom wires performed by Larade
  et al. \cite{22} pronounced a result consistent with that of Lang and Avouris.
  Analogous phenomenon was also observed in the metallic Na atom wires
  \cite{23}. Except for atom wires, the oligomer is another interesting molecular wire.
  For these oligomer wires, early theory presented a semi-empirical law called the
  conductance exponential law, $G=G_{m} e^{-\gamma L}$, here L is the length of
  the oligomer, $\gamma$ is a damping factor \cite{16}.
  This law is valid for the low bias voltage.
  Latterly, a full ab initio computation was applied to some kinds of
  oligomers and reached results conformed with this law \cite{19,20}.
  By using the Landauer formulation combined with the density functional theory (DFT)
  Tada et al.  \cite{21} studied several oligomer compounds and
  declared that the conductances for most of them conform with the exponential law, but there are
  two types of oligomers that disobey the law, polyphenanthrene dithiolates(PPh(n)DTs)
  and polyacene dithiolates(PA(n)DTs), with a conductance
  oscillation, similar to that in atom wires.
  The recent experiment of the oligothiophene dithiolates((n)T1DTs) shows an unusual
  length dependence of conductance for the bias lower than 1.2V.
  Although 4T1DT is longer than
  3T1DT, the conductance of 4T1DT is higher than 3T1DT's \cite{12}.
  The motivation of our study is to discover and understand the
  unusual length dependence of conductance for oligothiophene and other organic molecules.

  In this letter we employ a full ab initio method to calculate the transport behavior
  for three kinds
  of oligomers, (n)T1DTs, PPh(n)DTs and PA(n)DTs (see Fig.1).
  Our numerical results confirm the non-classical conductance length dependence
  of (n)T1DTs at low
  bias, which was shown in the experiment \cite{12}, and find a classical conductance
  length dependence
  coexisted at high bias. And the results of PPh(n)DTs and
  PA(n)DTs are far different from the semi-empirical calculation \cite{21}.
  Under low bias, both PPh(n)DTs and PA(n)DTs obey the exponential law qualitatively.
  With bias expanded into the high regime,
  PPh(n)DTs' length dependence retains the classical behavior,
  the longer the molecule is, the lower the conductance is. But for
  PA(n)DTs, the exponential law is broken down under high bias.
  In the high bias range the conductance will
  be inversely proportional to the molecule length,
  the long molecule wires have higher conductance than
  the short ones, which is opposite to the classical length dependence of conducting wires.
  Our computations show that coupling between the
  molecule and Au electrodes plays an important role. The
  coupling could shift and expand the molecule orbitals dramatically
  and change the HOMO-LUMO
  gap by different manners due to their distinct electronic structures.
  Coupled with Au leads, the HOMO-LUMO gap of
  PPh(n)DTs almost keeps a fixed value with the molecule length increased.
  On the contrary, the HOMO-LUMO gap of
  (n)T1DTs and PA(n)DTs will decrease obviously with increase in the molecule
  length. The distance between the Fermi level and HOMO or LUMO
  levels determines the conductance gap of the molecules and
  thus affects their length dependence of conductance.
  This point will be discussed latter in more details.

  After considerable effort was made by many people, theories based on the
  first-principles calculation have been built commendably. Here, we
  combine the density functional theory and nonequilibrium Green's function
  theory by means of the self-consistent procedure to solve the
  transport problem of the electrode-molecule-electrode open system \cite{4,5,6,7,8,9}.
  The advantage of such method is its theoretical rigor assured by the self-consistent
  procedure and its sensitivity to bonding and surface chemistry at
  contacts. The NEGF equations are as follows
\begin{equation}G^{R}=(E^{+}S-F-\Sigma^{R}_{1}-\Sigma^{R}_{2})^{-1},\end{equation}
\begin{equation}\label{2}\Sigma_{i}=\tau_{i}g_{i}\tau_{i}^{\dag},\ (i=1,2)  \end{equation}
\begin{equation}\rho=\int dE[G^{R}(f_{1}\Gamma_{1}+f_{2}\Gamma_{2})G^{A}/2\pi],\end{equation}
  where $S$ and $F$ are the overlap matrix and Fock matrix of the
  molecular part. $\Sigma^{R}_{1}$ ($\Sigma^{R}_{2}$) are the left (right) self
  energy. In our computation, the self-consistent procedure in Gaussian03 \cite{24} is
  extended from an isolated molecule to the lead-molecule-lead open system.
  The calculations for all the quantities in the open system, e.g.,
  the surface Green's functions
  $g_i$, the self-energies $\Sigma_i$, the broadening function $\Gamma_i$, and the density
  matrix $\rho$ are included in the main program of Gaussian03 as the subroutine programs.
  After this procedure is converged, transmission $T$ and current $I$ can be obtained
\begin{equation}T=Trace(\Gamma_{1}G\Gamma_{2}G^{\dag} ),\end{equation}
\begin{equation}I=(-2e/h)\int^{\infty}_{-\infty}
dET(E,V)(f_{1}(E)-f_{2}(E)).\end{equation}
  The computation detail is described in \cite{9}, where the DFT
  method with B3PW91 and LANL2DZ basis are adopted. The Fermi level of Au leads is set as $-5.1$ eV.

\begin{figure}
\includegraphics[scale=0.7,bb=60 32 320 558]{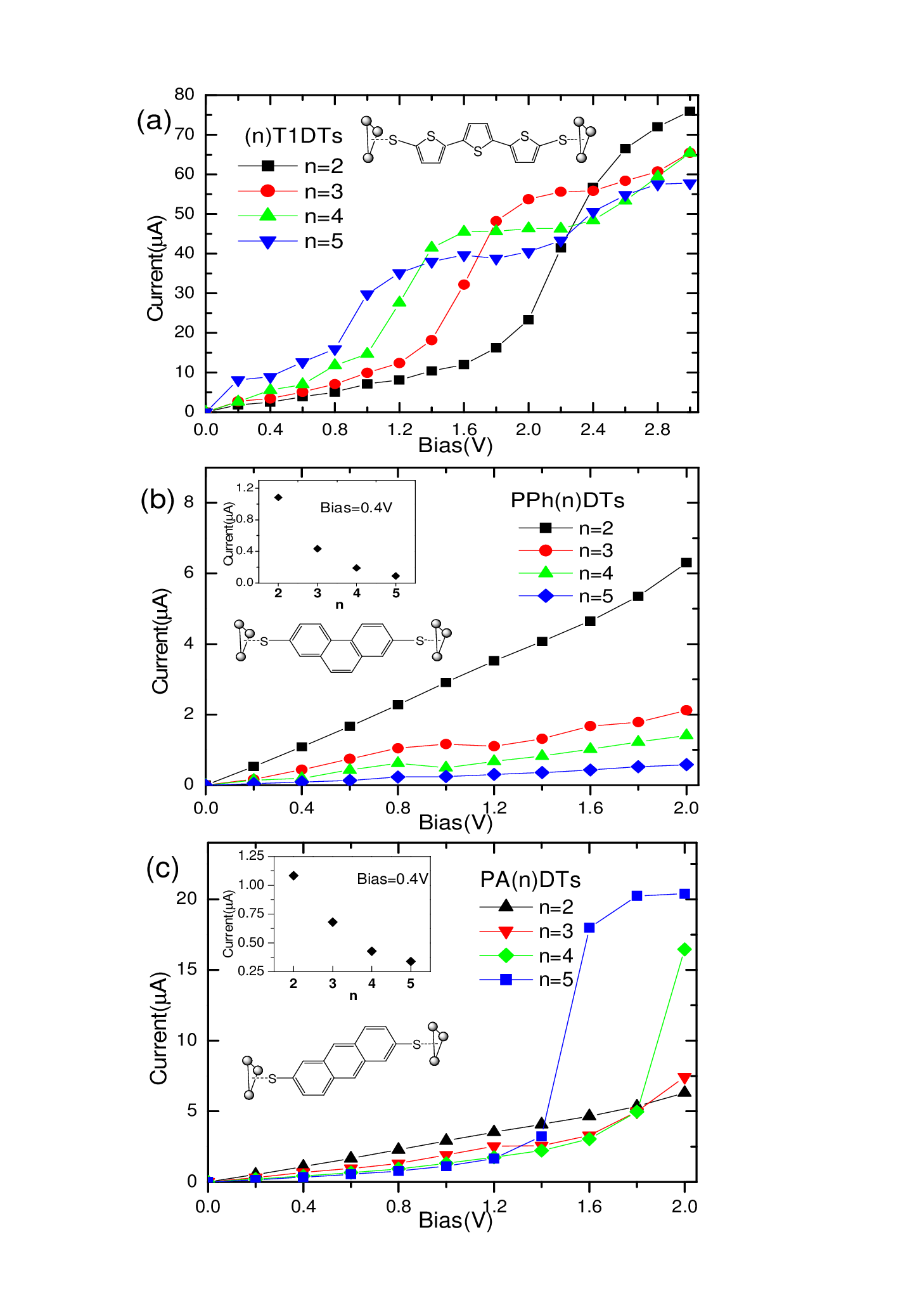}
\caption{I-V characteristics of the three oligomers. (a) (n)T1DTs
show the unusual length dependence in the low bias regime and the
classical length dependence at bias $V=3$ V. (b) PPh(n)DTs show a
classical length dependence. The inset illustrates that they obey
the Magoga's exponential law at low bias $V=0.4$ V. (c) PPh(n)DTs
present the classical length dependence at the low bias regime and
the quantum length dependence at bias $V=2$ V. The inset
illustrates that they obey the Magoga's exponential law at low
bias $V=0.4$ V. \label{fig1}}
\end{figure}

  The calculated I-V characteristics of (n)T1DTs, PPh(n)DTs and PA(n)DTs are
  shown in Fig.1, n is the number of thiophene ring (a) or benzene
  ring (b, c). (n)T1DTs with small conductance gap have bigger current than other
  two oligomers. Its length dependence of conductance is unusually non-classical under low bias,
  which is consistent well with the experiment reported by Xu
  et al. recently \cite{12}. They measured 3T1DT and 4T1DT in the bias
  regime from 0 V to 1 V and found that the conductance of 3T1DT is smaller than
  4T1DTs'. Not only does our calculation agree with their experiment result, but
  also it shows a very interesting return to the classical
  length dependence of conductance at high bias, near $V=3$ V.
  PPh(n)DTs illustrate the classical length dependence under the whole voltage range
  from $V=1$ V to $2$ V [Fig.1(b)],
  and the exponential law ($G=G_{m} e^{-\gamma L}$) qualitatively under low bias (inset of Fig.1(b)).
  Fig.1(c) shows a different conductance length dependence for PA(n)DTs.
  Under small bias voltage, the conductance exponential law
  is still satisfied (see inset of Fig.1(c)),
  but at high bias, the exponential law breaks down,
  some longer molecules have higher conductance than those shorter ones,
  which leads to an irregular conductance oscillation
  with respect to the n number. If bias increases
  continually to approach a special value, the classical length
  dependence of conductance will be completely reversed, that
  is, longer the molecule is, higher the conductance is. This
  special voltage is $2$ V, where the order of current values is
  PA(5)DTs, PA(4)DTs, PA(3)DTs, and PA(2)DTs, in a descending order.

  Transmission spectrum of the biased molecule is an
  effective tool to understand the I-V characteristics. It denotes the
  position of HOMO and LUMO peaks and the related distance between the Fermi energy and HOMO or
  LUMO, so one can determine the point where the first steep increase of
  current comes out.

\begin{figure}
\includegraphics[scale=0.7, bb=37 28 341 551]{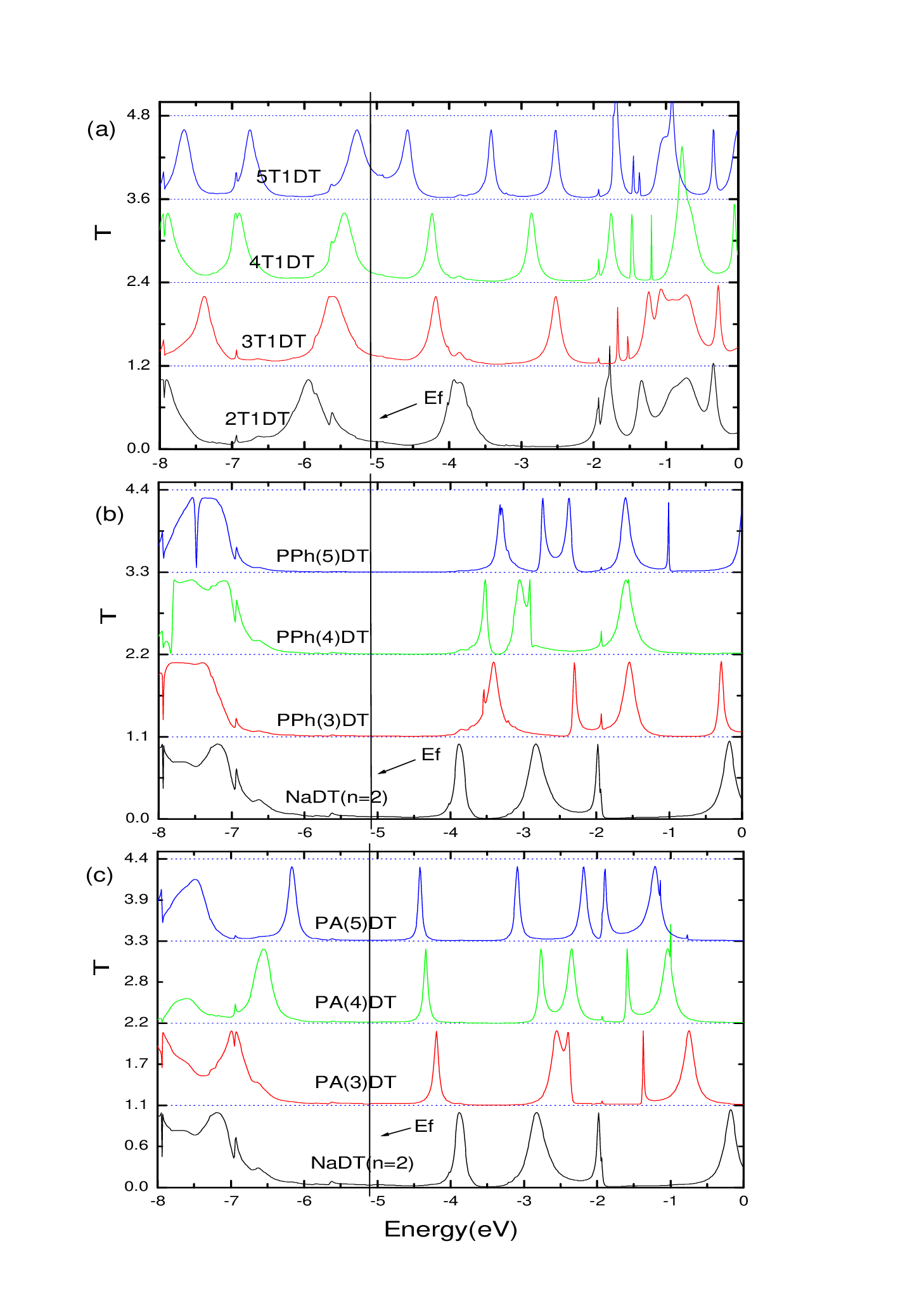}
\caption{ Transmission spectrum of the three oligomers under bias
0.2V (shifted vertically for visibility). (a) (n)T1DTs, the
HOMO-LUMO gap, which is small, and the distance between HOMO and
Fermi energy decrease with the rise of n number. (b) PPh(n)DTs,
the HOMO-LUMO gap, which is large, and the distance between LUMO
and Fermi energy increase with the rise of n number. (c) PA(n)DTs,
the HOMO-LUMO gap and the distance between LUMO and Fermi energy
decrease with the rise of n number. \label{fig2}}
\end{figure}

  As we know that the small HOMO-LUMO gap and the short distance
  between the LUMO (or HOMO) and the Fermi energy of the electrodes
  favor the higher conductivity.
  In Fig.2 HOMO-LUMO gaps of (n)T1DTs are very
  small and decrease obviously while number n increases.
  The peaks in the transmission spectrum or DOS spectrum illustrate
  the expanded molecular levels when the molecule is coupled with metal leads.
  Every expanded peak has tails overlapped each other which form
  the near-resonance channel. If HOMO peak and LUMO peak is very
  close, their strongly overlapped tail of transmission causes a big conduction value.
  Compared with other molecules, (n)T1DTs have very
  small HOMO-LUMO gaps with a great overlap in the low bias regime.
  This is the reason why (n)T1DTs have larger currents than
  other molecules under low bias. Furthermore, when number n
  increases, the HOMO-LUMO gap of (n)T1DTs decreases obviously.
  Hence, the overlap becomes stronger and the transmission in the low bias regime
  rises to induce an enhance of current. We can see clearly that
  the non-classical length dependence is just conduced by
  the drop of HOMO-LUMO gap and the distance between the HOMO and Fermi energy with increase of n number.

  Unlike (n)T1DTs, the HOMO-LUMO gap of PPh(n)DTs
  increases with the number of benzene rings, which indicates
  an ascending of the conductance gap with n number. LUMO, which is nearer Fermi energy than HOMO,
  dominates the position where
  the first resonance tunnelling happens and where the steep increase of current
  appears. The HOMO position of PPh(n)DTs is independent of n number.
  On the contrary, the HOMO-LUMO gap of PA(n)DTs decrease and the positions of
  HOMO and LUMO both approach to Fermi level when n
  number enhanced, which is similar to (n)T1DTs. Hence, the steep increase of current of
  long PA(n)DTs with small HOMO-LUMO gap will come out earlier
  than the short ones because its HOMO or LUMO resonance enters the bias window more
  rapidly than others. Under a little higher bias, a long
  PA(n)DT, e.g. PA(5)DT, throws out the resonance tunnelling and
  carries high current, meanwhile, the short molecule, e.g.
  PA(2)DT is still in the tunnel regime with low transmission
  coefficient. Therefor, the current of PA(5)DT is higher than that of PA(2)DT
  in the high bias regime. Because of such descending of HOMO-LUMO gap
  with n number, the I-V curves of PP(n)DTs will
  intercross, it is the reason why an irregular conductance oscillation
  and even the classical-reversed length dependence
  emerges in the high bias regime.

  At the same time, we notice that, in low bias regime, both PPh(n)DTs and PA(n)DTs
  have small current, which decreases exponentially with the rise of n number,
  because the low bias range is at the non-resonance tunnelling regime between HOMO
  and LUMO.
  In Fig.2(b) the contribution from the tail of LUMO to conduction becomes smaller while n number
  increases. In Fig.2(c) the situation is in reverse.
  The HOMO-LUMO gaps of PPh(n)DTs and PA(n)DTs are bigger than
  that of (n)T1DTs, which determines that the conductances of PPh(n)DTs and PA(n)DTs
  in low bias regime are mainly determined by the molecule length and thus
  exhibit the length dependence consistent with Magoga's exponential law.

\begin{figure}
\includegraphics[scale=0.4, angle=-90, bb=49 63 384 519]{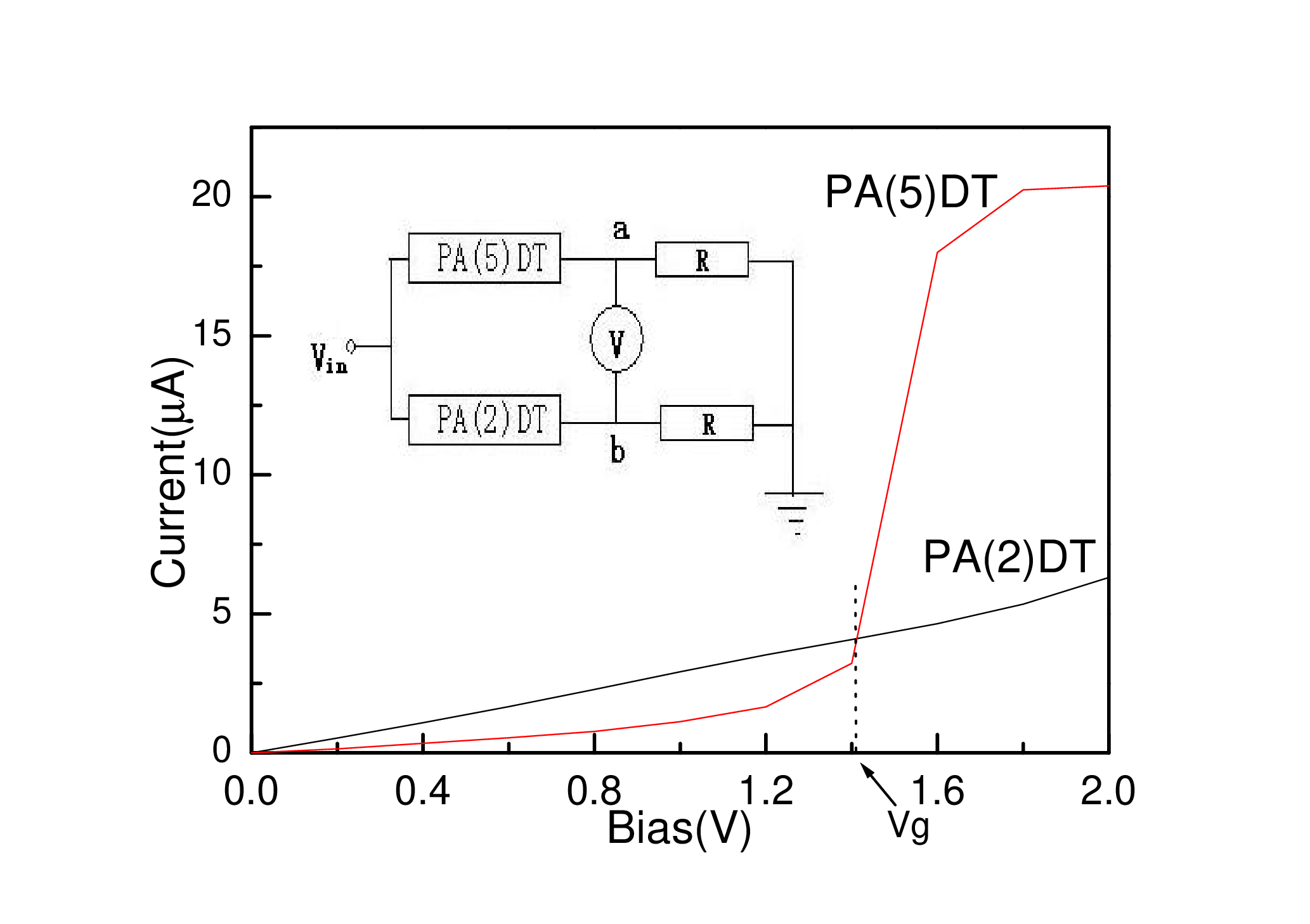}
\caption{ Molecule bridge with the rectifying property. The
resistances of PA(2)DT and PA(5)DT are controlled by the input
voltage $V_{in}$, which induce the sign variation in $V_{ab}$.
\label{fig3}}
\end{figure}

  The intercross of I-V curves of PA(5)DT and PA(2)DT can be
  employed for designing the molecular bridge as shown in Fig.3.
  $V_{g}$ denotes the intercross point. Below the point, the resistance of PA(5)DT is larger
  than PA(2)DT's, while above $V_g$ the situation will be in reverse.
  When the input voltage $V_{in}$ exceeds the intercross point,
  $V_{ab}$ changes from the positive to the negative one.

  In a summary, we employ the first-principles method based on DFT combined with the
  nonequilibrium Green's Function theory to study the transport
  property of (n)T1DTs, PPh(n)DTs and PA(n)DTs. Conductance of the three oligomers
  shows the different length dependence of conduction. PPh(n)DTs are usualy like as classical wires, but
  (n)T1DTs and PA(n)DTs demonstrate a crossover between the quantum conductance length dependence
  and the classical one controlled by the applied bias.
  The root reason is due to the fact that with the
  increase in the ring number in the oligomer compounds, HOMO and LUMO of the three oligomers
  approach to the different directions with respect to Fermi energy.
  The possible application of the length dependence of molecule conductance
  is proposed at last.

  We thank Dr. A. Ghosh, Prof. H. Guo and Prof. H.P. Cheng
for helpful discussions. This work is supported by the National
Science Foundation of China (NSFC) under Grant Nos. 90206031 and
10574024, and Special Coordination Funds of the Ministry of
Education, Culture, Sports, Science and Technology of the Japanese
Government. The author also would like to express their sincere
thanks to the support from the staff at the Center for
Computational Materials Science of IMR, Tohoku University for the
use of the SR8000 G/64 supercomputer facilities.

\end{document}